\documentclass{osa-article}

\journal{osajournal}%
\articletype{Research Article}
\usepackage{subfig}
\begin{document}
\title{Cross-Phase Modulation Instability in PM ANDi Fiber-Based Supercontinuum Generation}

\author{Etienne Genier\authormark{1,2,*}, Amar N. Ghosh\authormark{2}, Swetha Bobba\authormark{2},  Patrick Bowen\authormark{1}, Peter M. Moselund\authormark{1}, Ole Bang\authormark{1,3}, John M. Dudley\authormark{2}, and Thibaut Sylvestre\authormark{2}   }

\address{\authormark{1}NKT Photonics A/S, Blokken 84, DK-3460, Birker\o d, Denmark\\
\authormark{2}Institut FEMTO-ST, UMR 6174 CNRS-Universit\'{e} de Franche-Comt\'{e}, 25030 Besan\c{c}on, France\\
\authormark{3}DTU Fotonik, Department of Photonics Engineering, Technical University of Denmark, 2800 Kgs. \\ \hspace{0.15cm} Lyngby, Denmark }
\email{\authormark{*}etge@nktphotonics.com}

\begin{abstract}
We demonstrate broadband supercontinuum generation in an all-normal dispersion polarization-maintaining photonic crystal fiber and we report the observation of a cross-phase modulation instability sideband that is generated outside of the supercontinuum bandwidth. We demonstrate this sideband is polarized on the slow axis and can be suppressed by pumping on the fiber's fast axis. We theoretically confirm and model this nonlinear process using phase-matching conditions and numerical simulations, obtaining good agreement with the measured data.
\end{abstract}

    All-normal dispersion (ANDi) optical fibers have recently emerged as attractive platforms to improve the noise and coherence of supercontinuum (SC) generation beyond the limits of anomalous SC generation (SCG) [1-4]. ANDi SCG is based on two fully coherent nonlinear effects: self-phase modulation (SPM) and optical wave breaking (OWB) [2,5] while anomalous SCG is typically susceptible to or even generated by incoherent nonlinear effects [6]. Despite this, ANDi SCG has its own limitations, and being both very sensitive to Raman noise [2-4,7] and requiring low and flat fiber dispersion engineering that is technically challenging to achieve [1]. When pumping with femtosecond pulses, it has been shown that other factors should be considered including polarization modulation instability (PMI) or the amplitude noise of the laser, which both can drastically degrade the relative intensity noise (RIN) and coherence [3,4,8]. These factors limit the available parameter space for coherent SCG, however, fs-pumped ANDi SCG still has significant potential to generate temporally coherent SC with realistic laser parameters, a feature that is hard in the anomalous dispersion regime. This gives such systems potential in a range of fields including optical coherence tomography (OCT), optical metrology, photoacoustic imaging, and spectroscopy [9-12].  \\
\indent In this work, we investigate SCG in a PM-ANDi silica photonic crystal fiber (PCF) with a femtosecond stable optical parametric oscillator (OPO) with intention to suppress PMI. However, in doing this, we discovered the generation of a sideband outside the SC bandwidth which was not observed in previous PM-ANDi SCG [13,14]. We identify this sideband as the result of cross-phase modulation instability (XPMI) process that builds up from coherent SCG and OWB. As it is described in [15,16], XPMI is usually observed when a beam is launched at a 45 $^{\circ}$ angle from the principal axis of a highly birefringent fiber. This beam is then split into two linearly polarized modes on each axis that will nonlinearly interact with each other to generate two frequency-detuned and cross-polarized four-wave mixing (FWM) sidebands [15]. However, this XPMI process has never been observed before through the stimulation of a fs-SCG but only via spontaneous generation of the interaction of picosecond or nanosecond pulses.\\ \indent Our results show that we can generate a stimulated XPMI sideband in a PM-ANDi PCF using femtosecond pulses. As expected, this sideband is most powerful while pumping the fiber at 45$^{\circ}$ off the axes. We also demonstrate this sideband can be completely suppressed when pumping the fiber on the fast axis. Significantly, we note that while this sideband is observable outside the SC bandwidth with a low power pump, at higher powers the bandwidth of the supercontinuum will cover the sideband. \\
\indent The experimental setup used to observe and analyze SC and XPMI generation in the PM-ANDi PCF is shown in Fig. 1. As a pump laser, we used a Ti:Sa femtosecond pulsed laser (Coherent Chameleon) tunable from 680 nm - 1080 nm, delivering 200 fs pulse duration at a 80 MHz repetition rate with a maximum average power of 450 mW at 1040 nm.  The output power is controlled using a variable neutral density filter (ND). A half-wave plate is used to turn the input polarization state at the fiber input while the polarizer at the output of the fiber is used to observe the spectral content of the light of each axis. A 40x microscope objective is used to couple the light into the 40 cm of PM ANDi PCF -- the NL-1050-NE-PM from NKT Photonics. This fiber has a relative hole size of d/L = 0.45, a small hole-to-hole pitch of 1.44 $\mu$m, and a nonlinear coefficient of $\gamma$=26.8 W$^{-1}$km$^{-1}$ at 1040 nm. A set of 2 aspheric lenses is used to collimate the output beam and then focus it to the multimode pick-up fiber. 

\begin{figure*}[h!]
\centering
    \includegraphics[scale=0.5]{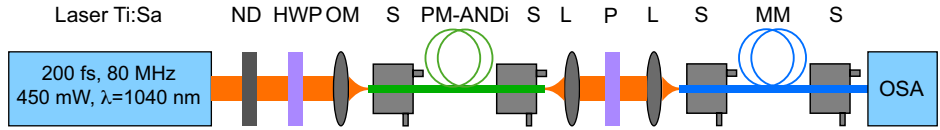}
    \caption{Schematic of the setup, including a wavelength tunable Ti:Sa femtosecond laser, a variable neutral density filter (ND), a half wave plate (HWP), a 40x microscope objective (OM), aspheric lenses (L), 3D translation stages (S), 40 cm of all normal dispersion PCF (PM-ANDi), a polarizer (P), 2 m of multimode pick-up fiber (MM), and an optical spectrum analyzer (OSA). \vspace{-0.5\baselineskip}}
\end{figure*}

\indent The dispersion was measured using white-light interferometry and calculated for an idealized structure with a uniform hole structure with fixed pitch and hole diameter using COMSOL. The COMSOL calculation fits the measurements well and is shown as the blue curve in Fig. 2. 

\begin{figure}[h!]
\vspace{-1\baselineskip}
\centering
    \includegraphics[scale=0.275]{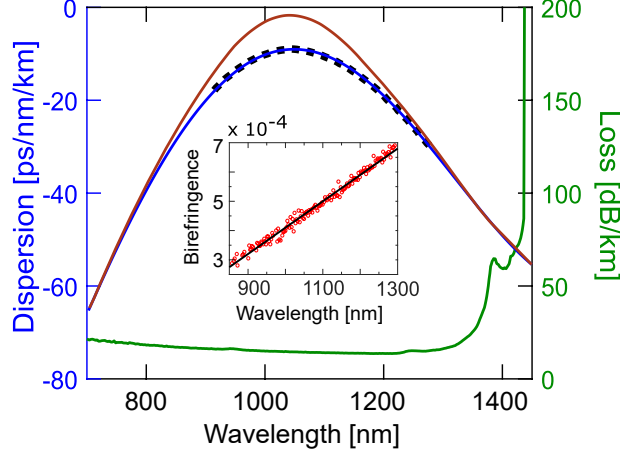}
    \caption{Comsol (solid blue), experimental (black dots) and modeled (dash brown) dispersion profiles, and fiber losses (solid green) of the NL-1050-NE-PM ANDi PCF. The inset shows the dispersion of the group birefringence: the linear fitting (solid black) and measured data (circles red).}
\end{figure}

The numerical modelling using the experimental dispersion data did not reproduce the experimental results, this is probably due to the uncertainties on this dispersion curve which increase when the dispersion is close to zero. We therefore considered the dispersion given by the brown curve, which is shifted upwards in the center as the dispersion approaches zero. This is still within the measurement uncertainties and is able to reproduce the experiments as we will demonstrate. The dispersion profile has a minimum of -13 ps/nm/km at 1040 nm and is rather symmetrical within the low-loss window. The polarization-maintaining effect of this fiber is stress-rod induced, with a slight degree of core-ellipticity that causes a linearly increasing birefringence [17], which goes from 2.5$.10^{-4}$ at 850 nm to 6.8$.10^{-4}$ at 1300 nm, as shown in Fig. 2 (red circles).

We pumped the fiber at 1040 nm, at the minimum dispersion wavelength (MDW), thus we should expect the broadest SC spectrum for a given power and input angle. Figure 3 shows the spectral evolution while pumping on the slow axis, fast axis and at 45$^{\circ}$.  The broadest SC spectrum (bandwidth at -20 dB is 460 nm) is obtained by pumping on the slow axis while the narrowest is obtained by pumping at 45$^{\circ}$. The spectrum generated when pumping on the fast axis is narrower than when pumping on the slow axis, which could be due to core ellipticity inducing a difference in the mode field diameter and thus in the nonlinear coefficient. 

\begin{figure}[h!]
\centering
    \includegraphics[scale=0.275]{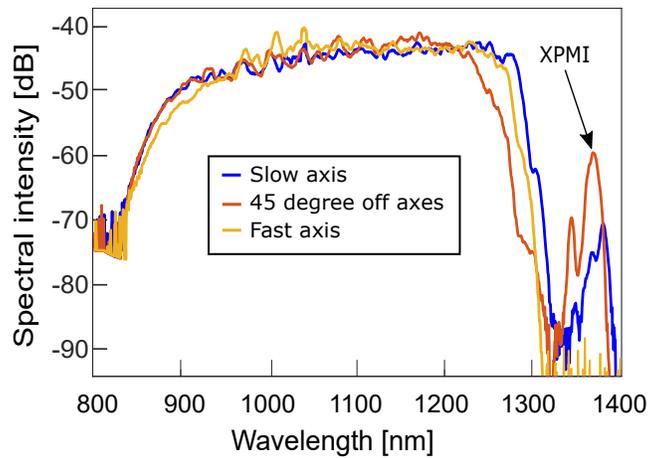}
    \caption{Experimental SC spectra for different input polarization with 220 mW output power.}
\end{figure} 

The SC spectrum obtained when pumping at 45$^{\circ}$ is narrower again because only half the power is available for spectral broadening in each axis and due to the temporal walk-off, which tends to eliminate the influence of cross-phase modulation [18].  Interestingly a sideband appears at 1360 nm, which is strongest when pumping at 45$^{\circ}$ and which we identify as XPMI. When pumping on the slow axis, the sideband intensity is reduced by 12 dB but is still clearly observable. Finally, when pumping on the fast axis the sideband appears to be completely suppressed with an extinction of at least 25 dB.
To understand the appearance of this sideband and the efficiencies regarding the input polarization angle, let us recall that the phase matching condition for XPMI should give rise to a Stokes sideband on the slow axis and an anti-Stokes sideband on the fast axis because of positive (normal) group-velocity dispersion (GVD) [6,19]. This means to stimulate the generation of an idler-pumped sideband in the Stokes side of the spectrum we need energy in the pump (on both axes) and energy in the anti-Stokes sideband (aligned to the fast axis).

Experimentally, we use a pump laser with polarization extinction ratio of 40 dB and inject the light into the  PM-ANDi fiber, whose input face was end-collapsed to remove back reflections into the Ti:Sa laser. At the output we measure a maximum achievable PER of 17 dB, which has a relatively even distribution over the whole SC bandwidth. This shows that, at some point in the propagation, light has deviated from the input axis to become distributed more over both axes, probably due to the collapse of the fiber holes. This can explain why it is possible to observe a Stokes sideband at 1360 nm even when the input beam is almost aligned on the slow axis, though the efficiency of this sideband will be much lower than when the beam is aligned at 45° of the axis, as shown in Fig. 3. 

\begin{figure}[h!]
\vspace{-0.\baselineskip}
\centering
    \includegraphics[scale=0.275]{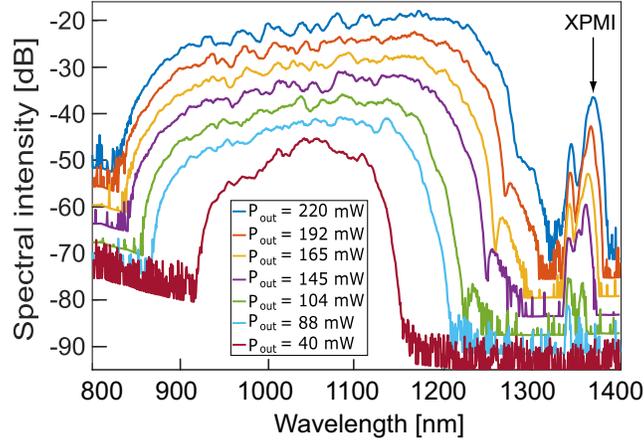}
    \caption{SC evolution for different pump power while pumping at 45$^{\circ}$ from the two axes (power offset for clarity). }
\end{figure}

Figure 4 shows the evolution of the SC spectrum as a function of the pump power for an input polarization at 45$^{\circ}$. The SC bandwidth is 430 nm at -20 dB (845 nm - 1275 nm) for an average output power of 220 mW. The XPMI sideband grows and slightly broadens to longer wavelengths when increasing the coupled power. It starts to appear only when SC extends past 1200 nm and when the OWB also starts to appear (See the green spectrum in Fig. 4). In addition, there is no observable anti-Stokes sideband outside the SC, even when observing the spectrum over a wide bandwidth (600 nm - 2000 nm). 

\begin{figure}[b!]
\centering
    \includegraphics[scale=0.275]{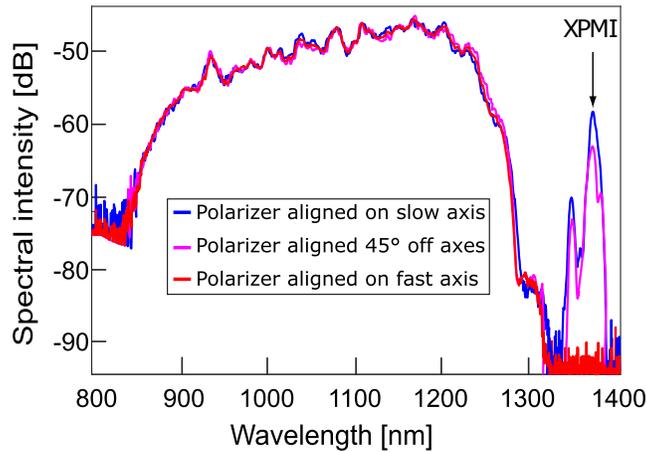}
    \caption{SC evolution for different polarizer orientation  while pumping at 45$^{\circ}$ from the two axes.}
\end{figure}

\indent Adding a polarizer at the output of the fiber to analyze  the sideband polarization angle, we show in Fig. 5 the SC spectrum measured after the polarizer as a function of the polarizer angle at maximum output power (220 mW average power). We can notice the 3 spectra obtained by aligning the polarizer on the slow axis (blue curve), fast axis (red curve) and at 45$^{\circ}$ of the axes (pink curve) have a similar bandwidth and shape. We observe the most powerful sideband when the polarizer is aligned on the slow axis and a 10 dB suppression when aligning the polarizer at 45$^{\circ}$ of the axes. Finally, aligning the polarizer on the fast axis totally suppresses the sideband, confirming that the sideband is polarized along the slow axis. 

\begin{figure}[h!]
\centering
    \includegraphics[width=8.5cm]{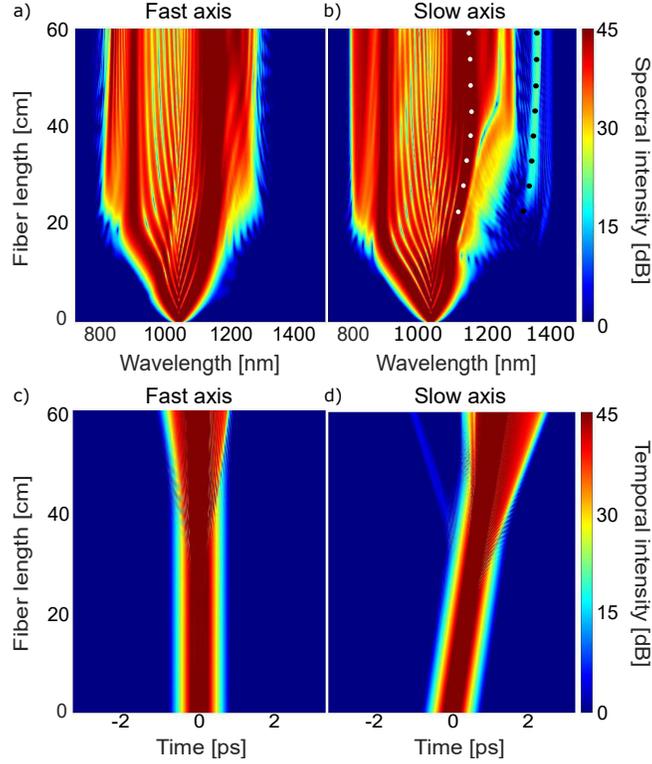}
    \caption{Simulated spectral (top) and temporal (bottom) SC evolution on each fiber axis as a function of fiber length. The dots represent the theoretical XPMI wavelength (black) using the redder SPM wavelength (white) as a pump. }
\end{figure}

To simulate the SCG in the PM-ANDi fiber, we use a Matlab code solving the two coupled generalized nonlinear Schrodinger equations (CGNLSE) for highly birefringent fibers as described in [20]. We used as input parameters a pump wavelength of 1040 nm, a sech-shaped pump pulse with duration of 200 fs (full width half maximum intensity), 12 kW peak power, a longer fiber length of 60 cm for better visibility, the loss profile as described in Fig. 2 (green curve) and the birefringence values taken from the inset. One-photon-per-mode noise and intensity noise of 1 \% was added to our input condition and the results were sampled average over 20 simulations.  Using these parameters, we obtain a quite good agreement between simulation and experimental results. This is shown in Figs. 6(a-d) that depict both the spectral and temporal intensity dynamics on each axis when pumping the PM-ANDi fiber at 45° . First the numerical SC bandwidth at -20 dB level is estimated to be 450 nm which is very close to the experimental one (430 nm). Second we can clearly see the generation of a small signal at 1360 nm polarized on the slow axis, which fits with the experimentally observed XPMI sideband (See Fig. 6b). We can also notice the sideband appears after 20 cm of propagation exactly where the OWB sets in and stops the red-shift of the SPM lobes. Interestingly, the temporal trace plotted in Fig. 6(d) reveals that the signal at 1360 nm behaves as a small dispersive wave (DW) shed by the pump pulse on the slow axis, in a way akin to the DW emission by OWB in the anomalous dispersion regime [21,22]. 

To get further into details, in Fig. 7, we plot the theoretical XPMI sidebands as a function of pump wavelength using the well-known formula for the XPMI frequency shift [23], \\
$\Omega_{s}(\lambda,z) = \frac{\delta n(\lambda)}{2 c \beta_{2}(\lambda) } \left( 1 + \sqrt{1-4\beta_{2}(\lambda) \gamma P_{0}(z) \left( \frac{c}{\delta n (\lambda)} \right)^{2} } \right)$, \\ where $\delta n$ is the group birefringence, $P_{0}$ the total peak power, and $\beta_{2}$ the the wavelength dependent GVD. The XPMI wavelengths are plotted in Fig. 7 for different peak powers of 12 kW, 6.5 kW, 4 kW, 2 kW and 0.  From the modelling, the slow axis XPMI generation is seen to be closely linked to the long wavelength SPM lobe in the slow axis, marked with white dots. According to this conjecture, the XPMI gain is only efficient when the red shift of the SPM lobe, acting as the pump, slows down, and for lengths not much longer  than the walk-off length of 22 cm. This is exactly what is observed: the XPMI peak is first growing after about 20cm when the red-shift of the SPM lobe is stopped by OWB, and after about 30 cm the power in the XPMI peak does not grow anymore.  Looking in more detail we find that the peak power decreases from 12 kW at z=0 to 1.5 kW at z=60 cm. From the corresponding phase-matching curves in Fig. 7 we see that the XPMI phase-matched to the center wavelength of the SPM lobe initially at 2787 nm ($P_{0}$=12 kW, $\lambda_{SPM}$=1040 nm), then it rapidly decreases because of the red-shift of the SPM lobe. The observed final SPM lobe wavelength of 1152 nm is seen to generate XPMI at 1360 nm in the valley of the linear  phase-matching curve, which nicely corresponds to the numerically and experimentally observed XPMI wavelength, for the adjusted dispersion profile. The observed SCG induced XPMI generation thus requires a delicate balance between  strong SPM stopped sufficiently before say twice the walk-off length by OWB.

\begin{figure}[t!]
\centering
   \includegraphics[scale=0.275]{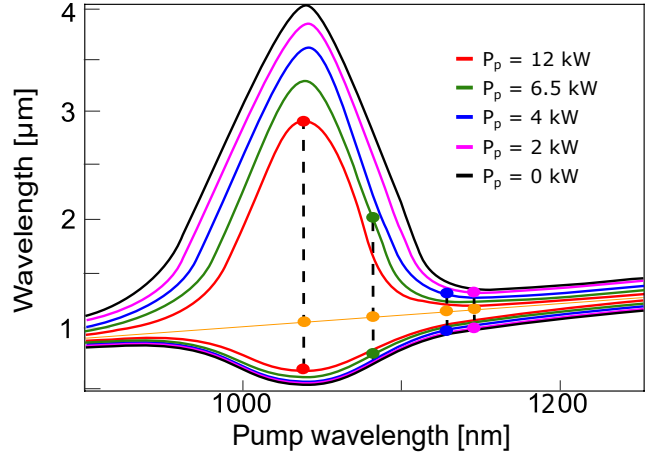}
  \caption{Phase-matching map as a function of total peak power, 12 kw (red), 6.5 kW (green), 4 kW (blue), 2 kW (pink), 0 (black). The orange dots on the orange straight line for the pump wavelength show the numerically observed red SPM wavelength (white dots in Fig. 6b) and the other colored dots mark the corresponding XPMI wavelength. }
\end{figure}

In conclusion, we have reported the observation of cross-phase modulation instability while pumping a PM-ANDi PCF with a femtosecond mode-locked laser. A sideband was generated through a XPMI process at 1360 nm during coherent supercontinuum generation from SPM and OWB. We demonstrated this sideband cannot be generated while pumping on the fast axis and is itself polarized along the fiber's slow axis. Further, clarifying that we were observing XPMI, theoretical calculation and simulation performed solving the CGNLSE confirmed the degenerate four wave mixing between a pump corresponding to the red edge of the SPM on the fiber's slow axis, an anti-Stokes idler pump in the central SPM area on the fiber fast's axis and a signal at 1360 nm polarized on the fiber's slow axis.  Our study is of substantial value to potential applications, such as OCT and metrology, which require ultra low-noise SC light sources. To achieve low-noise in these future SC sources, a high degree of suppression of XPMI will be required and for this a solid understanding of the underpinning physics. Indeed, this study shows that control the input polarization is very important to avoid noise amplification effect and thus keep the noise-free/stability given by a fs PM-ANDi SCG.\\

\noindent \textbf{Funding.} Horizon 2020 Framework Programme (722380); Agence Nationale de la Recherche (ANR) (ANR-15-IDEX-0003, ANR-17-EURE-0002). \\

\noindent \textbf{Disclosures.} The authors declare no conflicts of interest. \\

\noindent 1. Heidt, A. M., Hartung, A., Bosman, G. W., Krok, P., Rohwer, E. G., Schwoerer, H., \& Bartelt, H. (2011). Coherent octave spanning near-infrared and visible supercontinuum generation in all-normal dispersion photonic crystal fibers. Optics express, 19(4), 3775-3787. \\

\noindent 2. Heidt, A. M., Feehan, J. S., Price, J. H., \& Feurer, T. (2017). Limits of coherent supercontinuum generation in normal dispersion fibers. JOSA B, 34(4), 764-775. \\

\noindent 3. Gonzalo, I. B., Engelsholm, R. D., Sørensen, M. P., \& Bang, O. (2018). Polarization noise places severe constraints on coherence of all-normal dispersion femtosecond supercontinuum generation. Scientific reports, 8(1), 1-13. \\ 

\noindent 4. Genier, E., Bowen, P., Sylvestre, T., Dudley, J. M., Moselund, P., \& Bang, O. (2019). Amplitude noise and coherence degradation of femtosecond supercontinuum generation in all-normal-dispersion fibers. JOSA B, 36(2), A161-A167. \\

\noindent 5. Finot, C., Kibler, B., Provost, L., \& Wabnitz, S. (2008). Beneficial impact of wave-breaking for coherent continuum formation in normally dispersive nonlinear fibers. JOSA B, 25(11), 1938-1948. \\

\noindent 6. Dudley, J. M., Genty, G., \& Coen, S. (2006). Supercontinuum generation in photonic crystal fiber. Reviews of modern physics, 78(4), 1135. \\

\noindent 7. Møller, U., \& Bang, O. (2013). Intensity noise in normal-pumped picosecond supercontinuum generation, where higher-order Raman lines cross into anomalous dispersion regime. Electronics letters, 49(1), 63-65. \\

\noindent 8. Loredo-Trejo, A., López-Diéguez, Y., Velázquez-Ibarra, L., Díez, A., Silvestre, E., Estudillo-Ayala, J. M., \& Andrés, M. V. (2019). Polarization Modulation Instability in All-Normal Dispersion Microstructured Optical Fibers With Quasi-Continuous Pump. IEEE Photonics Journal, 11(5), 1-8. \\ 

\noindent 9. Dasa, M. K., Markos, C., Maria, M., Petersen, C. R., Moselund, P. M., \& Bang, O. (2018). High-pulse energy supercontinuum laser for high-resolution spectroscopic photoacoustic imaging of lipids in the 1650-1850 nm region. Biomedical optics express, 9(4), 1762-1770. \\

\noindent 10. Petersen, C. R., Prtljaga, N., Farries, M., Ward, J., Napier, B., Lloyd, G. R., ... \& Bang, O. (2018). Mid-infrared multispectral tissue imaging using a chalcogenide fiber supercontinuum source. Optics letters, 43(5), 999-1002. \\

\noindent 11. Israelsen, N. M., Maria, M., Mogensen, M., Bojesen, S., Jensen, M., Haedersdal, M., ... \& Bang, O. (2018). The value of ultrahigh resolution OCT in dermatology-delineating the dermo-epidermal junction, capillaries in the dermal papillae and vellus hairs. Biomedical optics express, 9(5), 2240-2265. \\

\noindent 12. Povazay, B., Bizheva, K., Unterhuber, A., Hermann, B., Sattmann, H., Fercher, A. F., ... \& Russell, P. S. J. (2002). Submicrometer axial resolution optical coherence tomography. Optics letters, 27(20), 1800-1802. \\ 

\noindent 13. Tarnowski, K., Martynkien, T., Mergo, P., Poturaj, K., Anuszkiewicz, A., Béjot, P., ... \& Urbanczyk, W. (2017). Polarized all-normal dispersion supercontinuum reaching 2.5 µm generated in a birefringent microstructured silica fiber. Optics express, 25(22), 27452-27463. \\ 

\noindent 14. Tarnowski, K., Martynkien, T., Mergo, P., Sotor, J., \& Soboń, G. (2019). Compact all-fiber source of coherent linearly polarized octave-spanning supercontinuum based on normal dispersion silica fiber. Scientific reports, 9(1), 1-8. \\ 

\noindent 15. Kudlinski, A., Bendahmane, A., Labat, D., Virally, S., Murray, R. T., Kelleher, E. J. R., \& Mussot, A. (2013). Simultaneous scalar and cross-phase modulation instabilities in highly birefringent photonic crystal fiber. Optics express, 21(7), 8437-8443. \\

\noindent 16. Drummond, P. D., Kennedy, T. A. B., Dudley, J. M., Leonhardt, R., \& Harvey, J. D. (1990). Cross-phase modulational instability in high-birefringence fibers. Optics communications, 78(2), 137-142. \\

\noindent 17. Hlubina, P., Ciprian, D., \& Kadulová, M. (2008). Wide spectral range measurement of modal birefringence in polarization-maintaining fibres. Measurement Science and Technology, 20(2), 025301. \\

\noindent 18. Sylvestre, T., Maillotte, H., Lantz, E., \& Gindre, D. (1997). Combined spectral effects of pulse walk-off and degenerate cross-phase modulation in birefringent fibers. Journal of Nonlinear Optical Physics \& Materials, 6(03), 313-320. \\

\noindent 19. Chen, J. S., Wong, G. K., Murdoch, S. G., Kruhlak, R. J., Leonhardt, R., Harvey, J. D., ... \& Knight, J. C. (2006). Cross-phase modulation instability in photonic crystal fibers. Optics letters, 31(7), 873-875. \\

\noindent 20. Ghosh, A. N., Meneghetti, M., Petersen, C. R., Bang, O., Brilland, L., Venck, S., ... \& Sylvestre, T. (2019). Chalcogenide-glass polarization-maintaining photonic crystal fiber for mid-infrared supercontinuum generation. Journal of Physics: Photonics, 1(4), 044003. \\

\noindent 21. Conforti, M., \& Trillo, S. (2013). Dispersive wave emission from wave breaking. Optics letters, 38(19), 3815-3818. \\ 

\noindent 22. Webb, K. E., Xu, Y. Q., Erkintalo, M., \& Murdoch, S. G. (2013). Generalized dispersive wave emission in nonlinear fiber optics. Optics letters, 38(2), 151-153. \\ 

\noindent 23. Agrawal, G. P. (2000). Nonlinear fiber optics. In Nonlinear Science at the Dawn of the 21st Century (pp. 195-211). Springer, Berlin, Heidelberg.

\end{document}